\documentclass[10pt, twocolumn, final]{IEEEtran}
\usepackage{amsfonts}
\usepackage{graphicx,cite}
\usepackage{amsmath}
\usepackage{amssymb}
\usepackage{verbatim}
\newtheorem{thm}{Theorem}
\begin{document}
\title{On the Capacity Equivalence with Side Information at Transmitter and
Receiver}
\author{Yong Peng and Dinesh Rajan, \emph{Senior Member, IEEE}
\thanks{The authors are with the Wireless Networking Lab, Dept. of Electrical Engineering,
Southern Methodist University, Dallas, TX 75275, USA
(\{ypeng,rajand\}@engr.smu.edu). This work has been supported in
part by the National Science Foundation through grant CCF 0546519.}}
\date{}
\maketitle

\begin{abstract}
In this paper, a channel that is contaminated by two independent
Gaussian noises $S\sim \mathcal{N}(0,Q)$ and
${Z_0}\sim\mathcal{N}(0,N_0)$ is considered. The capacity of this
channel is computed when independent noisy versions of ${S}$ are
known to the transmitter and/or receiver. It is shown that the
channel capacity is greater then the capacity when ${S}$ is
completely unknown, but is less then the capacity when ${S}$ is
perfectly known at the transmitter or receiver. For example, if
there is one noisy version of ${S}$ known at the transmitter only,
the capacity is $\frac{1}{2}\log(1+\frac{P}{Q(N_1/(Q+N_1))+N_0})$,
where $P$ is the input power constraint and $N_1$ is the power of
the noise corrupting $S$. Further, it is shown that the capacity
with knowledge of any independent noisy versions of $S$ at the
transmitter is equal to the capacity with knowledge of the
statistically equivalent noisy versions of $S$ at the receiver.
\end{abstract}
\section{INTRODUCTION}
Consider a channel in which the received signal, $\mathbf{Y}$ is
corrupted by two independent additive white Gaussian noise (AWGN)
sequences,~$\mathbf{S}\sim \mathcal{N}(0,Q\mathbf{I}_n)$ and
$\mathbf{Z}_0\sim\mathcal{N}(0,N_0\mathbf{I}_n)$, where
$\mathbf{I}_n$ is the identity matrix of size~$n$. The received
signal is of the form,
\begin{equation}
\mathbf{Y} = \mathbf{X} + \mathbf{S} + \mathbf{Z}_0,
\label{eq:chan_model}
\end{equation}
where $\mathbf{X}$ is the transmitted sequence for $n$ uses of the
channel. Let the transmitter and receiver each has knowledge of
independent noisy observations of one of the noises, $\mathbf{S}$.
We quantify the benefit of this additional knowledge by computing
the capacity of the channel in~(\ref{eq:chan_model}) and presenting
the coding scheme that achieves capacity. Our result indicates that
the capacity is of the form $\frac{1}{2} \log (1+\frac{P}{\mu Q +
N_0})$, where $0\le \mu \le 1$ is the residual fraction~(explicitly
characterized in Section~\ref{sec:model}) of the interference power,
$Q$, that can not be canceled with the noisy observations at the
transmitter and receiver.

One special case of our model is when a noisy version of
$\mathbf{S}$ is known only to the transmitter; our result in this
case is a generalization of Costa's celebrated result~\cite{Costa}.
In~\cite{Costa}, it is shown that the achievable rate when the noise
$\mathbf{S}$ is perfectly known at the transmitter is equivalent to
the rate when $\mathbf{S}$ is known at the receiver, and this rate
does not depend on the variance of $\mathbf{S}$. A new coding
strategy to achieve this capacity was also introduced
in~\cite{Costa} and is popularly referred to as dirty paper
coding~(DPC). We generalize Costa's result to the case of noisy
interference knowledge. We show that the capacity with knowledge of
a noisy version of $\mathbf{S}$ at the transmitter is equal to the
capacity with knowledge of a statistically equivalent noisy version
of $\mathbf{S}$ at the receiver. However, unlike~\cite{Costa} where
the capacity does not depend on the variance of~$\mathbf{S}$, in the
general noisy side information case, the capacity decreases as the
variance of~$\mathbf{S}$ increases.

We also compute the capacity when multiple independent noisy
observations of~$\mathbf{S}$ are available at the transmitter and
receiver. We show that in this case, it is sufficient for the
transmitter and receiver to each compute maximum-likelihood
estimates of~$\mathbf{S}$ based on their observations and then use
these estimates to achieve capacity using the coding strategy
proposed in Section~\ref{sec:Achieve}. Further, it is shown that the
capacity of a Gaussian channel with multiple independent
observations of~$\mathbf{S}$ known at the transmitter is equal
to the capacity with statistically similar observations known at the receiver.

The proposed model can have several potential applications. For
instance, consider the following scenario. Node~A is transmitting to
node~B, but whose signal is also received at nodes~C and~D who are
communicating with each other. Thus, nodes~C and~D can use the noisy
estimate of user~A's signal to improve the rate at which they
communicate.

In~\cite{Costa}, Costa adopted the random coding argument given by
Gel'fand and Pinsker~\cite{Gelfand} and El Gamal and
Heegard~\cite{Gamal}. In \cite{Gelfand}, the capacity of a discrete
memoryless channel with side information at the encoder is derived
and the result has been extended to the case when each of the
encoder and decoder has one of two correlated side information
\cite{Cover1}. Based on the channel capacity
$C=\max_{p(u,x|s)}\{I(U;Y)-I(U,S)\}$ given in \cite{Gelfand,Gamal},
Costa constructed the auxiliary variable~${U}$ as a linear
combination of ${X}\sim \mathcal{N}(0,P)$ and ${S}\sim
\mathcal{N}(0,Q)$ and showed that this simple construction of~${U}$
achieves capacity. In our proof for achievability, we also follow
the arguments from \cite{Gelfand,Gamal}. Further, similar
to~\cite{Costa}, the optimal auxiliary variable in our case is also
a linear combination of ${X}$ and the noisy observations of ${S}$ at
the encoder. Thus, our coding scheme can be viewed as a variation of
DPC.

Following the pioneering work of Costa, several extensions of DPC
have been studied, \emph{e.g.}, colored Gaussian noise~\cite{Yu},
arbitrary distributions of~$\mathbf{S}$~\cite{Cohen} and
deterministic sequences~\cite{Erez}. In \cite{Licks}, DPC has been
applied to the AWGN and jitter channel and the rate loss due to
imperfect synchronization at the decoder has been evaluated. DPC has
also been extensively applied to watermarking or information
embedding~\cite{Chen, Cohen1} applications and for the Gaussian
broadcast channel~\cite{Caire,Yu1}. In~\cite{Mazzotti} the authors
consider the case when $\mathbf{S}$ is perfectly known to the
encoder and a noisy version is known to the decoder. Most of the
analysis in~\cite{Mazzotti} is for discrete memoryless channels and
also causal knowledge of interference. The only result
in~\cite{Mazzotti} for Gaussian channel shows that there is no
additional gain due to the presence of the noisy estimate at the
decoder, since perfect knowledge is available at the encoder and a
DPC can be constructed. In contrast, in this paper we study the case
when only noisy knowledge of~$\mathbf{S}$ is available at both
transmitter and receiver.

Throughout this paper we only consider the case of non-causal
knowledge of the interference. Causal extensions to these should be
studied in future work. The rest of this paper is organized as
follows: Section~\ref{sec:model} introduces the system model and
also gives the main capacity result. Extensions of the result to
single and multiple noisy observations are evaluated in
Section~\ref{sec:generalization}. Brief concluding remarks are given
in Section~\ref{sec:conclusion}.

\section{SYSTEM MODEL, CAPACITY and ACHIEVABILITY \label{sec:model}}
In this section, we first introduce the channel model and then
compute the capacity of this channel.
\subsection{Channel Model}
\begin{figure}[htp]
\begin{center}
\includegraphics[width=3.4in]{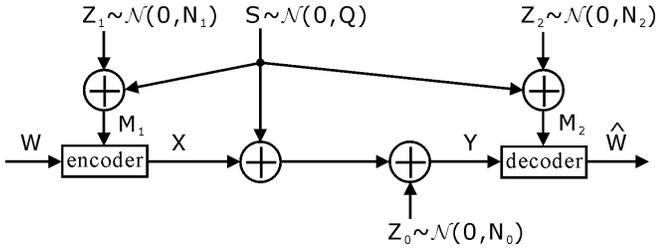}
\caption{Channel model.} \label{fig:DPC_1}
\end{center}
\end{figure}
The channel model is depicted in Fig. \ref{fig:DPC_1}. The
communication problem is to send an index, $W\in \{1,2,\ldots,K\}$,
to the receiver in $n$ uses of the channel. The transmission rate is
$R = \frac{1}{n}\log_2K$ bits per transmission. The output of the
channel is given in~(\ref{eq:chan_model}) and is contaminated by two
independent AWGN sequences, $\mathbf{S}\sim
\mathcal{N}(0,Q\mathbf{I}_n)$ and
$\mathbf{Z}_0\sim\mathcal{N}(0,N_0\mathbf{I}_n)$. Side information
$\mathbf{M}_1=\mathbf{S}+\mathbf{Z}_1$, which is noisy observations
of the interference is assumed to be available at the transmitter
for all $n$ uses of the channel. Similarly, noisy side information
$\mathbf{M}_2=\mathbf{S}+\mathbf{Z}_2$, is assumed to be available
at the receiver for all $n$ uses of the channel. The noise vectors
are distributed as $\mathbf{Z}_1 \sim \mathcal{N}(0,N_1
\mathbf{I}_n)$ and $\mathbf{Z}_2 \sim \mathcal{N}(0,N_2
\mathbf{I}_n)$.

Based on the index $W$ and the observation $\mathbf{M_1}$, the
encoder transmits one codeword, $\mathbf{X}$, from a $(2^{nR},n)$
code book. The codeword $\mathbf{X}$ must satisfy an average power
constraint of the form, $\frac{1}{n}\|\mathbf{X}\|^2 \leq P$. Let
$\hat{W}$ be the estimate at the receiver of the transmitted index;
an error occurs if $\hat{W}\neq W$.

\subsection{Channel Capacity}\label{sec:capacity}
\begin{thm}
Consider a channel of the form~(\ref{eq:chan_model}) with an average
transmit power constraint~$P$. Let independent noisy observations
$\mathbf{M}_1=\mathbf{S}+\mathbf{Z}_1$
and~$\mathbf{M}_2=\mathbf{S}+\mathbf{Z}_2$ of the
interference~$\mathbf{S}$ be available, respectively, at the
transmitter and receiver. The noise vectors have the following
distributions: $\mathbf{Z}_i \sim \mathcal{N}(0,N_i \mathbf{I}_n)$,
$i=0,1,2$ and $\mathbf{S} \sim \mathcal{N}(0,Q \mathbf{I}_n)$. The
capacity of this channel equals $\frac{1}{2}\log\left(1+\frac{P}{\mu
Q+N_0}\right)$, where $0\le
\mu=\frac{1}{1+\frac{Q}{N_1}+\frac{Q}{N_2}} \le 1$.
\end{thm}

\emph{Remark:} It is clear that $\mu= 0$ when either $N_1=0$ or
$N_2=0$. Consequently, the capacity is
$\frac{1}{2}\log\left(1+\frac{P}{N_0}\right)$, which is consistent
with the result in~\cite{Costa}\footnote{Costa's result is a special
case with $N_1=0$ and $N_2=\infty$.}. Further, $\mu=1$ when
$N_1\rightarrow \infty$ and $N_2\rightarrow \infty$, and the
capacity is $\frac{1}{2}\log\left(1+\frac{P}{Q+N_0}\right)$, which
is the capacity of a Gaussian channel with noise variance $Q+N_0$.
Thus, one can interpret~$\mu$ as the residual fractional power of
the interference that cannot be canceled by the noisy observations
at the transmitter and receiver.

\emph{Proof:} We first compute an outer bound on the capacity of
this channel. It is clear that the channel capacity can not exceed
$\max_{p(x|m_1,m_2)}I(X;Y|M_1,M_2)$, which is the capacity when both
$M_1$ and $M_2$ are known at the transmitter and receiver. Thus, a
capacity bound of the channel can be calculated as
\begin{align}
&~I(X;Y|M_1,M_2) = I(X;Y,M_1,M_2) - I(X;M_1,M_2) \nonumber \\
\leq&~ I(X;Y,M_1,M_2)\label{eq:cap_ub_step} \\
=&~H(X)+H(Y,M_1,M_2)-H(X,Y,M_1,M_2)\nonumber \\
=&~{\frac{1}{2}\log(2\pi e)^4P\left|\begin{array}{ccc}
P+Q+N_0 & Q  & Q\\
Q & Q+N_1 & Q\\
Q & Q & Q+N_2
\end{array} \right|}\nonumber\\
&~-{\frac{1}{2}\log(2\pi e)^4\left|\begin{array}{cccc}
P & P & 0 & 0\\
P & P+Q+N_0 & Q  & Q\\
0 & Q & Q+N_1 & Q\\
0 & Q & Q & Q+N_2\\
\end{array} \right|} \nonumber\\
=&~\frac{1}{2}\log\left(1+\frac{P}{\mu Q+N_0}\right).
\label{eq:capacity}
\end{align}
where $\mu=\frac{1}{1+\frac{Q}{N_1}+\frac{Q}{N_2}}$. Note that the
inequality in~(\ref{eq:cap_ub_step}) is actually a strict equality
since $I(X;M_1,M_2)=0$.

\subsection{Achievability of Capacity}\label{sec:Achieve}
We now prove that the capacity computed in (\ref{eq:capacity}) is
achievable. The codebook generation and encoding method we use
follows the principles introduced in~\cite{Gelfand, Gamal}. The
construction of auxiliary variable is similar to \cite{Costa}.

\emph{Random codebook generation:}
\begin{enumerate}
\item Generate $2^{nI(U;Y,M_2)}$ i.i.d. length-$n$ codewords
$\mathbf{U}$, whose elements are drawn i.i.d. according to
$U\sim\mathcal{N}(0,P+\alpha^2(Q+N_1))$, where $\alpha$ is a
coefficient to be optimized.
\item Randomly place the $2^{nI(U;Y,M_2)}$ codewords
$\mathbf{U}$ into $2^{nR}$ cells in such a way that each of the
cells has the same number of codewords. The codewords and their
assignments to the $2^{nR}$ cells are revealed to both the
transmitter and the receiver.
\end{enumerate}

\emph{Encoding:}
\begin{enumerate}
\item Given an index $W$ and an observation, $\mathbf{M_1}=\mathbf{M_1}(i)$,
of the Gaussian noise sequence, $\mathbf{S}$, the encoder searches
among all the codewords $\mathbf{U}$ in the $W^{th}$ cell to find a
codeword that is jointly typical with $\mathbf{M_1}(i)$. According
to the joint asymptotic equipartition property (AEP)~\cite{Cover},
it is easy to show that if the number of codewords in each cell is
larger than or equal to $2^{nI(U,M_1)}$, the probability of finding
such a codeword $\mathbf{U}=\mathbf{U}(i)$ exponentially approaches
$1$ as $n\rightarrow \infty$.
\item Once a jointly typical pair
$(\mathbf{U}(i),\mathbf{M_1}(i))$ is found, the encoder calculates
the codeword to be transmitted as
$\mathbf{X}(i)=\mathbf{U}(i)-\alpha \mathbf{M_1}(i)$. With high
probability, $\mathbf{X}(i)$ will be a typical sequence which
satisfies $\frac{1}{n}\|\mathbf{X}(i)\|^2\leq P$.
\end{enumerate}

\emph{Decoding:}
\begin{enumerate}
\item Given $\mathbf{X}(i)$ is transmitted, the received signal is
$\mathbf{Y}(i)=\mathbf{X}(i)+\mathbf{S}+\mathbf{Z_0}$. The decoder
searches among all $2^{nI(U;Y,M_2)}$ codewords $\mathbf{U}$ for a
sequence that is jointly typical with $\mathbf{Y}(i)$. By joint AEP,
the decoder will find $\mathbf{U}(i)$ as the only jointly typical
codeword with probability approaching 1.
\item Based on the knowledge of the codeword assignment to
the cells, the decoder estimates $\hat{W}$ as the index of the
cell that $\mathbf{U}(i)$ belongs to.
\end{enumerate}

\emph{Proof of achievability:}

 Let $U=X+\alpha M_1=X+\alpha
(S+Z_1)$, $Y=X+S+Z_0$ and $M_2=S+Z_2$, where $X\sim
\mathcal{N}(0,P)$, $S\sim \mathcal{N}(0,Q)$ and $Z_i\sim
\mathcal{N}(0,N_i),~i=0,1,2$ are independent Gaussian random
variables. To ensure that with high probability, in each of the
$2^{nR}$ cells, at least one jointly typical pair of $\mathbf{U}$
and $\mathbf{M_1}$ can be found. The rate,~$R$, which is a function
of $\alpha$, must satisfy
\begin{equation}
R(\alpha)\leq I(U;Y,M_2)-I(U;M_1).\label{eq:rate}
\end{equation}
The two mutual informations in~(\ref{eq:rate}) can be calculated
respectively as
\begin{align}
&~I(U;Y,M_2) \nonumber\\
=&~H(U)+H(Y,M_2)-H(U,Y,M_2)\nonumber\\
=&~\frac{1}{2}\log 2\pi
e\left[P+\alpha^2(Q+N_1)\right]\nonumber\\
&~+\frac{1}{2}\log(2\pi e)^2\left|\begin{array}{cc}
P+Q+N_0 & Q \\
Q & Q+N_2
\end{array}\right|\nonumber\\
&~-\frac{1}{2}\log(2\pi e)^3\cdot \nonumber\\
&~~~~\left|\begin{array}{ccc}
P+\alpha^2(Q+N_1) & P+\alpha Q  & \alpha Q\\
P+\alpha Q & P+Q+N_0 & Q\\
\alpha Q & Q & Q+N_2
\end{array}\right|\nonumber\\
=&~\frac{1}{2}\log\left(\frac{\left[P+\alpha^2(Q+N_1)\right]\left|\begin{array}{cc}
P+Q+N_0 & Q \\
Q & Q+N_2
\end{array}\right|}{\left|\begin{array}{ccc}
P+\alpha^2(Q+N_1) & P+\alpha Q  & \alpha Q\\
P+\alpha Q & P+Q+N_0 & Q\\
\alpha Q & Q & Q+N_2
\end{array}\right|}\right)\label{eq:I1}
\end{align}
and
\begin{equation}
I(U;M_1)=\frac{1}{2}\log\left(\frac{P+\alpha^2(Q+N_1)}{P}\right).\label{eq:I2}
\end{equation}
Substituting~(\ref{eq:I1}) and~(\ref{eq:I2}) into~(\ref{eq:rate}), we find
\begin{align}
&R(\alpha) \le \frac{1}{2} \log P[(Q+P+N_0)(Q+N_2)-Q^2 ]
\nonumber\\
&- \frac{1}{2}\log
\left\{\alpha^2[Q(P+N_0)(N_1+N_2)+(Q+P+N_0)N_1N_2]\right.\nonumber\\
&\left.-2\alpha QPN_2+P(QN_0+QN_2+N_0N_2) \right\}. \label{eq:rate1}
\end{align}
We can now find the optimal coefficient $\alpha^*$ that maximizes
the right hand side of~(\ref{eq:rate1}) using the extreme
value theorem. After simple algebraic
manipulations, the optimal coefficient is computed as
\begin{equation}
\alpha^*=
\frac{QP{{N_2}}}{Q(P+N_0)({N_1}+{N_2})+(Q+P+N_0){N_1N_2}}.\label{eq:alpha}
\end{equation}
Substituting for $\alpha^*$ into~(\ref{eq:rate}), the maximal rate
is found to be
\begin{equation}
R(\alpha^*)=\frac{1}{2}\log\left(1+\frac{P}{\mu
Q+N_0}\right)\label{eq:achieve}
\end{equation}
with $\mu=\frac{1}{1+\frac{Q}{N_1}+\frac{Q}{N_2}}$, which is exactly
the upper bound on capacity~(\ref{eq:capacity}).

\section{SPECIALIZATION AND GENERALIZATION}\label{sec:generalization}
The coding scheme to achieve capacity in Section \ref{sec:Achieve}
can be easily specialized to the case when a single observation is
available at either the encoder or decoder. Further, it can be
generalized to the scenario when multiple independent observations
are available at the encoder and decoder.
\subsection{Single Observation}
\subsubsection{Noisy estimate at receiver only}

\begin{figure}[h]
\begin{center}
\includegraphics[width=3.4in]{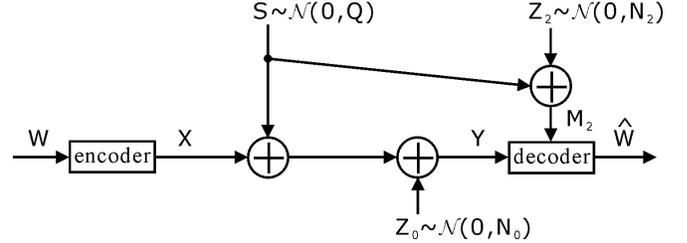}
\caption{Channel model with an observation of $\mathbf{S}$ at the
receiver.} \label{fig:DPC_2}
\end{center}
\end{figure}

Fig. \ref{fig:DPC_2} shows the channel model when the observation of
$\mathbf{S}$ is only available at the receiver. This channel is
equivalent to our original model when $N_1\rightarrow \infty$. The
capacity of the channel is given by
\begin{equation}
I(X;Y,M_2)=\frac{1}{2}\log\left(1+\frac{P}{Q\left(\frac{N_2}{Q+N_2}\right)+N_0}\right).
\label{eq:C_rx}
\end{equation}

\subsubsection{Noisy estimate at transmitter only: Generalization
of Dirty Paper Coding}

\begin{figure}[h]
\begin{center}
\includegraphics[width=3.4in]{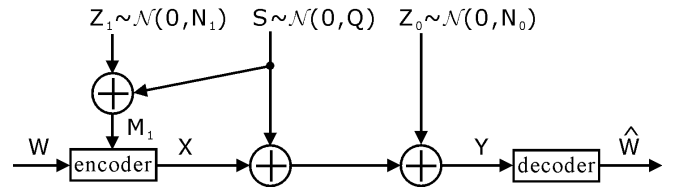}
\caption{Channel model with an observation of $\mathbf{S}$ at
transmitter.} \label{fig:DPC_3}
\end{center}
\end{figure}

Fig. \ref{fig:DPC_3} shows the channel model when the observation of
$\mathbf{S}$ is only available at the transmitter. This is a special
case of our original model with $N_2\rightarrow \infty$. The
capacity of the channel is
\begin{eqnarray}
I(X;Y|M_1)=\frac{1}{2}\log\left(1+\frac{P}{Q\left(\frac{N_1}{Q+N_1}\right)+N_0}\right).\label{eq:capacity_1}
\end{eqnarray}
The achievability can be easily shown following the same steps as in
Section \ref{sec:Achieve} with $N_2\rightarrow \infty$. Note that
when $N_1=0$, the channel model further reduces to Costa's dirty
paper coding channel model~\cite{Costa}.

\emph{Remarks:}

1) In~\cite{Costa}, Costa showed the surprising fact that the
channel capacity with additive Gaussian interference known to the
transmitter only is the same as the channel capacity with the
interference known to the receiver only. This paper extends that
result to the case of noisy interference. Indeed, by setting
$N_1=N_2$ in (\ref{eq:C_rx}) and (\ref{eq:capacity_1}), we can see
that the capacity with noisy interference known to transmitter only
equals the capacity with a statistically similar noisy interference
known to receiver only.

2) From~(\ref{eq:capacity_1}), one may intuitively interpret the
effect of knowledge of~$M_1$ at the transmitter. Indeed, a fraction
$\frac{Q}{Q+N_1}$ of the interfering power can be canceled using the
proposed coding scheme. The remaining $\frac{N_1}{Q+N_1}$ fraction
of the interfering power, $Q$, is treated as `residual' noise. Thus,
unlike Costa's result~\cite{Costa}, the capacity in this case
depends on the power~$Q$ of the interfering source: For a
fixed~$N_1$, as~$Q$ increases, the capacity decreases. The capacity
approaches $\frac{1}{2}\log\left(1+\frac{P}{N_1+N_0}\right)$ with
$Q\rightarrow \infty$.

\subsection{Multiple Independent Observations}

\begin{figure}[htp]
\begin{center}
\includegraphics[width=3.4in]{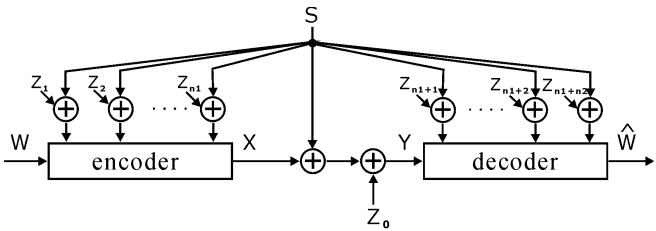}
\caption{Channel model with multiple independent observations of
$\mathbf{S}$.} \label{fig:DPC_n}
\end{center}
\end{figure}
Let there be $n_1$ independent observations
$\mathbf{M}_{1},\mathbf{M}_2, \ldots,$ $\mathbf{M}_{n_1}$ of
$\mathbf{S}$ at the transmitter and $n_2$ independent observations
$\mathbf{M}_{n_1+1}, \mathbf{M}_{n_1+2},\ldots,\mathbf{M}_{n_1+n_2}$
at the receiver, as shown in Fig.~\ref{fig:DPC_n}. Similar to
Section~\ref{sec:capacity}, an upper bound on the capacity of this
channel can be computed as
\begin{align}
&~I(X;Y|M_1,M_2,\ldots,M_{n_1+n_2})\nonumber\\
\leq&~ I(X;Y,M_1,M_2,\ldots,M_{n_1+n_2})\nonumber \\
=&~H(X)+H(Y,M_1,M_2,\ldots,M_{n_1+n_2})\quad\quad\quad~\quad\quad\quad\nonumber\\
&-H(X,Y,M_1,M_2,\ldots,M_{n_1+n_2})\nonumber
\end{align}
\begin{align}
=&~\frac{1}{2}\log(2\pi e)^{n_1+n_2+2}\cdot \nonumber \\
&~P\left|\begin{array}{cccc}
P+Q+N_0 & Q & \cdots & Q\\
Q & Q+N_1 & \ddots & \vdots\\
\vdots & \ddots &\ddots & Q\\
Q & \cdots & Q & Q+N_{n_1+n_2}
\end{array} \right|\nonumber\\
&-\frac{1}{2}\log(2\pi e)^{n_1+n_2+2}\cdot \nonumber \\
&~\left|\begin{array}{ccccc}
P & P & 0 & \cdots & 0\\
P & P+Q+N_0 & Q & \cdots & Q\\
0 & Q & Q+N_1 & \ddots & \vdots\\
\vdots & \vdots & \ddots &\ddots & Q\\
0 & Q & \cdots & Q & Q+N_{n_1+n_2}
\end{array} \right| \nonumber\\
=&~\frac{1}{2}\log\left(1+\frac{P}{\hat{\mu}Q+N_0}\right)\label{eq:gen_capacity}
\end{align}
where $\hat{\mu}=\frac{1}{1+\frac{Q}{N_1}+\frac{Q}{N_2}+\cdots
+\frac{Q}{N_{n_1+n_2}}}$ and $N_1,N_2,\ldots,N_{n_1+n_2}$ are the
variances of the Gaussian noise variables,
$Z_1,Z_2,\ldots,Z_{n_1+n_2}$, corresponding to the $n_1+n_2$
observations.

\emph{Achievability of capacity:} To show the achievability, the
coding process follows the same random coding argument as in Section
\ref{sec:Achieve}. The main difference is that in this case, we
first construct one estimate of the interference at the
transmitter~(similarly at the receiver) based on the multiple noisy
observations and then use this estimate in the coding process. Thus,
we omit the detailed development and show only the main steps.

At the transmitter, upon receipt of the $n_1$ independent
observations, the encoder makes a maximum likelihood estimation
(MLE) of $\mathbf{S}$, which is given as
\begin{equation}
\hat{S}_{1}=\arg{\underset{S} \max}f_{S}(M_1,M_2,\ldots,M_{n_1}|S)
\end{equation}
where
\begin{equation}
f_{S}(M_1,M_2,\ldots,M_{n_1}|S)=\prod_{l=1}^{n_1}\frac{1}{\sqrt{2\pi
N_l}}\exp\left(-\frac{(M_l-S)^2}{2N_l}\right). \label{eq:dist}
\end{equation}
Taking logarithm on both side of (\ref{eq:dist}) and differentiating
with respect to $S$, the estimate, $\hat{S}_{1}$, which maximizes
$f_S$ is found to be
\begin{equation}
\hat{S}_{1}=\sum_{l=1}^{n_1}\frac{M_l}{
\sum_{k=1}^{n_1}\frac{N_l}{N_k}}=S+\sum_{l=1}^{n_1}\frac{Z_l}{\sum_{k=1}^{n_1}\frac{N_l}{N_k}}.
\label{eq:S_MLE}
\end{equation}
Thus, the variance of the estimation error at the transmitter can be
computed as
\begin{equation}
\hat{N}_1=\text{Var}\left(\sum_{l=1}^{n_1}\frac{Z_l}{\sum_{k=1}^{n_1}\frac{N_l}{N_k}}\right)=
\frac{1}{\sum_{l=1}^{n_1}\frac{1}{N_l}}.\label{eq:N_1}
\end{equation}
Similarly, at the receiver, the decoder also computes the MLE,
$\hat{S}_{2}$ of $S$ based on the $n_2$ independent observations.
The variance of the estimation error at the receiver is given by
\begin{equation}
\hat{N}_2=\frac{1}{\sum_{l=n_1+1}^{n_1+n_2}\frac{1}{N_l}}.\label{eq:N_2}
\end{equation}
Thus, using MLE, the multiple observations is equivalent to one
observation each at the transmitter and receiver with estimation
noise variance $\hat{N}_1$ and $\hat{N}_2$. Essentially, the
construction of the auxiliary variable~${U}$ is similar to that in
Section~\ref{sec:Achieve} using $\hat{N}_1$ and $\hat{N}_2$,
\emph{i.e.}, ${U}=X+\alpha \hat{S}_{1}$. The achievable rate, $R^*$,
can then be found by substituting $\hat{N}_1$ and $\hat{N}_2$
instead of $N_1$ and $N_2$, respectively, in (\ref{eq:achieve}) and
is given by,
\begin{equation}
R^*=\frac{1}{2}\log\left(1+\frac{P}{\hat{\mu}Q+N_0}\right)\label{eq:gen_R1}
\end{equation}
where $\hat{\mu} =
\frac{1}{1+\frac{Q}{\hat{N}_1}+\frac{Q}{\hat{N}_2}}=\frac{1}{1+\frac{Q}{N_1}+\frac{Q}{N_2}+\cdots
+\frac{Q}{N_{n_1+n_2}}}$. Clearly, $R^*$ is the same as the upper
bound on capacity given in (\ref{eq:gen_capacity}).

\emph{Remark:} It is easy to see that the capacity expression is
symmetric in the noise variances at the transmitter and receiver. In
other words, having all the $n_1+n_2$ observations at the
transmitter would result in the same capacity. Thus, the
observations of $\mathbf{S}$ made at the transmitter and the
receiver are equivalent in achievable rate, as long as the
corrupting Gaussian noises have the same statistics. Further, if
there is an extra independent observation of $\mathbf{S}$ with
corrupting noise variance $N_{n_1+n_2+1}$, the fraction $\hat{\mu}$
in (\ref{eq:gen_R1}) decreases to
$\frac{1}{1+\frac{Q}{N_1}+\frac{Q}{N_2}+\cdots
+\frac{Q}{N_{n_1+n_2}}+\frac{Q}{N_{n_1+n_2+1}}}$. Thus, the
irremovable part of the noise power $Q$ decreases and the achievable
rate increases. It is clear that it does not matter whether this
extra observation is obtained at the transmitter or the receiver.

\section{CONCLUSION}\label{sec:conclusion}

In this paper, we derived the capacity region for a Gaussian noise
channel with additive Gaussian interference, when noisy estimates of
the interference are known at the transmitter and receiver. Our
results indicate that knowledge of the interference at the
transmitter gives the same capacity as knowledge of statistically
similar interference at the receiver. As noted earlier, all results
in this paper are derived assuming non-causal knowledge of the noisy
interference. Studying capacity with causal knowledge of the
interference should be investigated in future work.

\bibliographystyle{ieeetr}
\bibliography{DPC}
\end{document}